\documentclass[12pt]{article} 

\usepackage{ol2}
\usepackage{amsmath}

\begin{document}
\title{Homodyne locking of a squeezer}

\author{M. Heurs$^{1,3}$, I. R. Petersen$^1$, M. R. James$^2$, E. H. Huntington$^{3,*}$}
\affiliation{
$^1$School of Engineering and Information
Technology, University College, The
University of New South Wales, Canberra, ACT, 2600\\
$^2$Department of Engineering, Faculty of Engineering and Information Technology, The
Australian National University, Canberra, ACT, 0200\\
$^3$Centre for Quantum Computer Technology, School of Engineering and Information
Technology, University College, The
University of New South Wales, Canberra, ACT, 2600\\
$^*$Corresponding author: e.huntington@adfa.edu.au
}


\begin{abstract}

We report on the successful implementation of a new approach to locking the frequencies of an OPO-based squeezed-vacuum source and its driving laser.  The technique allows the simultaneous measurement of the phase-shifts induced by a cavity, which may be used for the purposes of frequency-locking, as well as the simultaneous measurement of the sub-quantum-noise-limited (sub-QNL) phase quadrature output of the OPO.  The homodyne locking technique is cheap, easy to implement and has the distinct advantage that subsequent homodyne measurements are automatically phase-locked.  The homodyne locking technique is also unique in that it is a sub-QNL frequency discriminator.

\end{abstract}


Nonclassical states are considered key resources in quantum information processing applications \cite{Nielsen00}, quantum key distribution systems \cite{Cerf07} and high-precision
measurements such as gravitational wave interferometry \cite{Chickarmane94}.  One particular nonclassical state for which a number of applications have been proposed is the vacuum-squeezed state, which is commonly produced via a sub-threshold optical parametric oscillator (OPO) in continuous-wave systems  \cite{Cerf07}.

In
such systems, a squeezed vacuum is created via the non-linear interaction between a classical pump field and the modes of a cavity \cite{BR04}.  Mechanical disturbances to the cavity make it impossible to create, measure or use the squeezed vacuum if the OPO cavity is left uncontrolled.  Consequently, all such systems are frequency-locked to the driving
laser.

In order to frequency-lock the OPO and driving laser together, an error signal must be generated that is bipolar and proportional to their frequency detuning.  Common approaches to extracting an appropriate error signal include the Pound-Drever-Hall (PDH) modulation-based technique \cite{pound83} or modulation-free techniques such as H\"{a}nsch-Couillaud \cite{hansch80} and tilt-locking \cite{shaddock99}.  These are all quite different techniques,
but they share in common the idea that they are implemented separately from the homodyne measurements of the squeezed cavity mode.  Such an approach leaves the squeezed
output mode undisturbed and ready to direct towards subsequent experiments or measurement apparatus.

Here we report on the successful implementation of an alternative scheme, which we call homodyne locking, that allows simultaneous measurement of
the phase-shift induced by the cavity (for the purposes of frequency locking) as well as phase-quadrature homodyne detection of the squeezed
output of the OPO.  The homodyne locking technique makes use of the principle that a non-resonant mode, which provides the phase reference for
frequency-locking, can also be used as the local oscillator in a homodyne detector.

The homodyne locking technique has the distinct advantage that any subsequent homodyne detector is automatically locked to a known phase angle.  This locking technique is easy and cost-effective to implement as well as, uniquely, being a sub-quantum-noise-limited frequency discriminator.

The homodyne locking technique is based on the observation
that measurements of the phase quadrature of the output of a cavity can be used to estimate the frequency detuning between the input to the cavity
and the cavity mode of interest \cite{CDC}.  That is, assuming a singly-resonant cavity, no pump depletion, and parametric de-amplification, the sub-threshold OPO is described by the
following set of equations \cite{BR04}:
\begin{eqnarray}
\dot a &=& -(\kappa + i \Delta)a - \chi a^{\dagger}+ \sqrt{2\kappa_{s}}A_{s} + \sqrt{2\kappa_{l}}\delta A_{l} \nonumber\\
A_{sqz}&=&\sqrt{2\kappa_{s}} a - A_{s} 
\label{cavity-1}
\end{eqnarray}
Here, $a$ denotes the annihilation operator for the squeezed cavity mode defined in an appropriate rotating frame.   The detuning
$\Delta$ represents the frequency deviation of the cavity's resonant frequency from the nominal laser frequency $\omega_0$.  The cavity decay
rates are determined by mirror reflectivities $R_i$ and are given by $\kappa_{i}=(1-R_{i})/{2\tau}$ for a cavity round trip time of $\tau$. The
total decay rate for the fundamental is $\kappa=\kappa_l+\kappa_{s}$. The seed field incident on the front face of the cavity is given by
${A}_{s}$ and the squeezed output field is $A_{sqz}$.   The amplitude and phase quadratures of any field $A_i$ are denoted $X_i^{+} = A_i+A_i^{\dagger}$ and
$X_i^{-}=iA_i-iA_i^{\dagger}$ respectively.

The intra-cavity annihilation operator can be decomposed into $a=\langle a \rangle +\delta{a}$ as can the seed $A_{s}=\langle
A_{s}\rangle+\delta A_{s}$.  Eq. \ref{cavity-1}, can be solved in steady-state to find that
\begin{equation}
X_{sqz}^{\pm} =\frac{{2\kappa_{s}}}{\kappa^2- \chi^2+\Delta^2}\left[ (\kappa \pm \chi)X_{s}^{\pm} \mp \Delta X_{s}^{\mp}\right] - X_{s}^{\pm}.
\label{alpha}
\end{equation}
From here on, we will assume without loss of generality that the input field is real.  Linearising Eq.
\ref{alpha} near $\Delta=0$, we find that steady-state measurements of the phase quadrature of the output field are linearly proportional to the
detuning and hence can be used as an error signal.

Eq. \ref{cavity-1} can also be solved in frequency space for the fluctuating terms to yield an expression for the squeezing spectrum:

\begin{equation}
\delta X_{sqz}^{\pm} =\frac{\delta X_{s}^{\pm} \left[ 2\kappa_{s}- (\kappa+ i\omega) \pm \chi\right] + \delta X_l^{\pm} 2\sqrt{\kappa_{s}\kappa_l}}{ (\kappa+i\omega) \mp \chi}
\label{squeezingSimple}
\end{equation}

Here we have set $\Delta=0$ and $\omega$ is a frequency measured relative to one of the longitudinal resonances of the OPO cavity.  We assume that $\omega$ is small compared to the free-spectral-range (FSR) of the OPO cavity.

Therefore, a single measurement of the phase quadrature of
the output field affords both a direct measurement of the squeezed output of the sub-threshold OPO, Eq. \ref{squeezingSimple}, and the detuning between the
cavity resonance and the input field as required for frequency locking, Eq. \ref{alpha}.

\begin{figure}[htbp]
  \centering
  \vspace{-2pt}
  \includegraphics[width=\linewidth,viewport=70 250 700 500,clip]{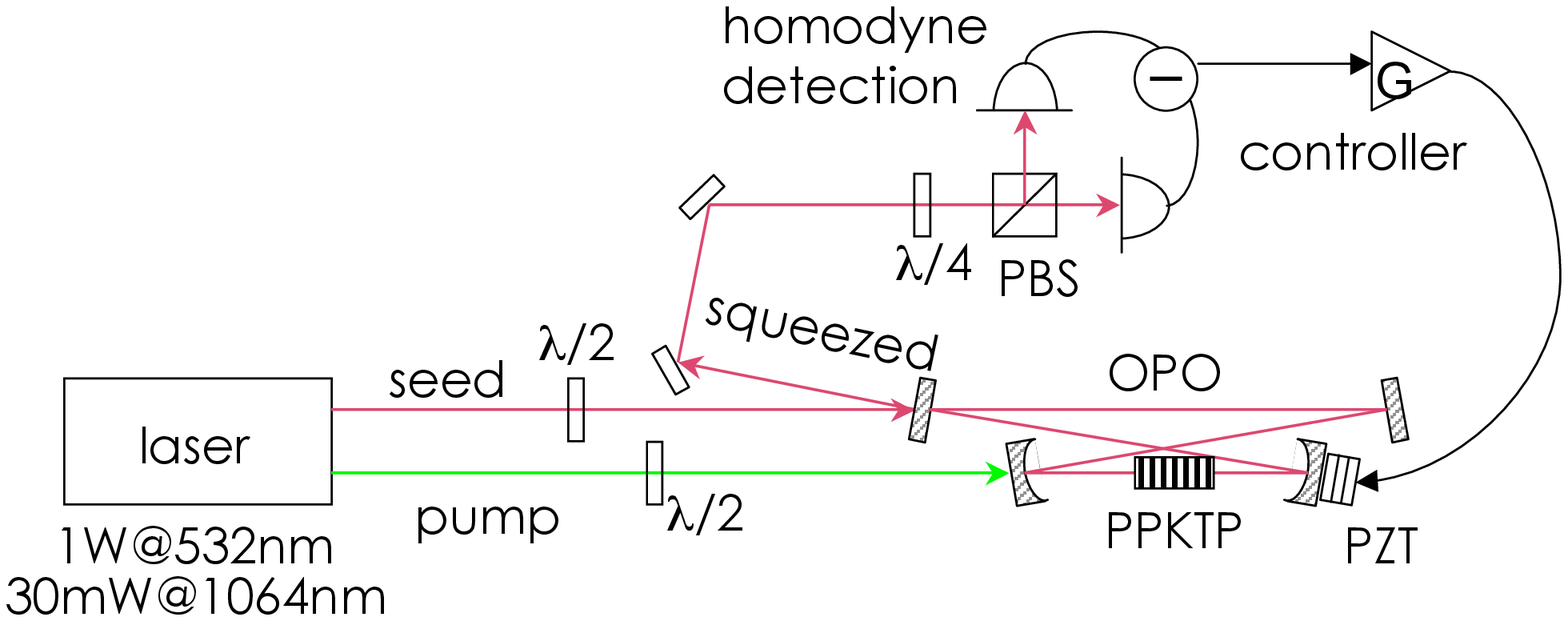}
  \caption{Schematic of homodyne locking experiment.}
  \label{schem}
\end{figure}

Fig. \ref{schem} illustrates schematically the experiment used to demonstrate homodyne locking.  The 532nm, frequency-doubled output of a
diode-pumped, miniature monolithic Nd:YAG laser is used to pump a sub-threshold OPO.  The nonlinear crystal is periodically-poled KTP, with a
phase-matching temperature of $33.5^{\circ}$C.  The OPO has a free-spectral range of 199
MHz and is operated with a parametric amplification of 3.9 dB and de-amplification of 2.6dB.  A small fraction of the source laser power is tapped
off prior to frequency-doubling and is used as the seed to the OPO and the local oscillator for the homodyne measurements

Quadrature homodyne measurement requires a bright, coherently related local oscillator (LO) with which to interfere the field of interest \cite{BR04}.  The phase of the LO relative to the field of interest determines the measurement quadrature.  In the experiments described here the LO
co-propagates with the seed to ensure that phase fluctuations are common-mode and hence that the  homodyne detector is passively locked to measure
the phase quadrature as required by Eqs. \ref{alpha} and \ref{squeezingSimple}.  

In the experiment described here 1\% of the power of the input field is set to the polarisation appropriate to the nonlinear
crystal in the OPO cavity.  The remaining 99\% of the power is in the orthogonal, non-resonant polarisation and acts as the LO.  This is not the
only means of implementing passive locking of the homodyne detector. In a previous experiment involving an empty cavity, the same effect was
achieved by using a spatially offset, but co-propagating LO \cite{qcmc}.

\begin{figure}[htbp]
  \centering
  \vspace{-2pt}
  \includegraphics[width=\linewidth]{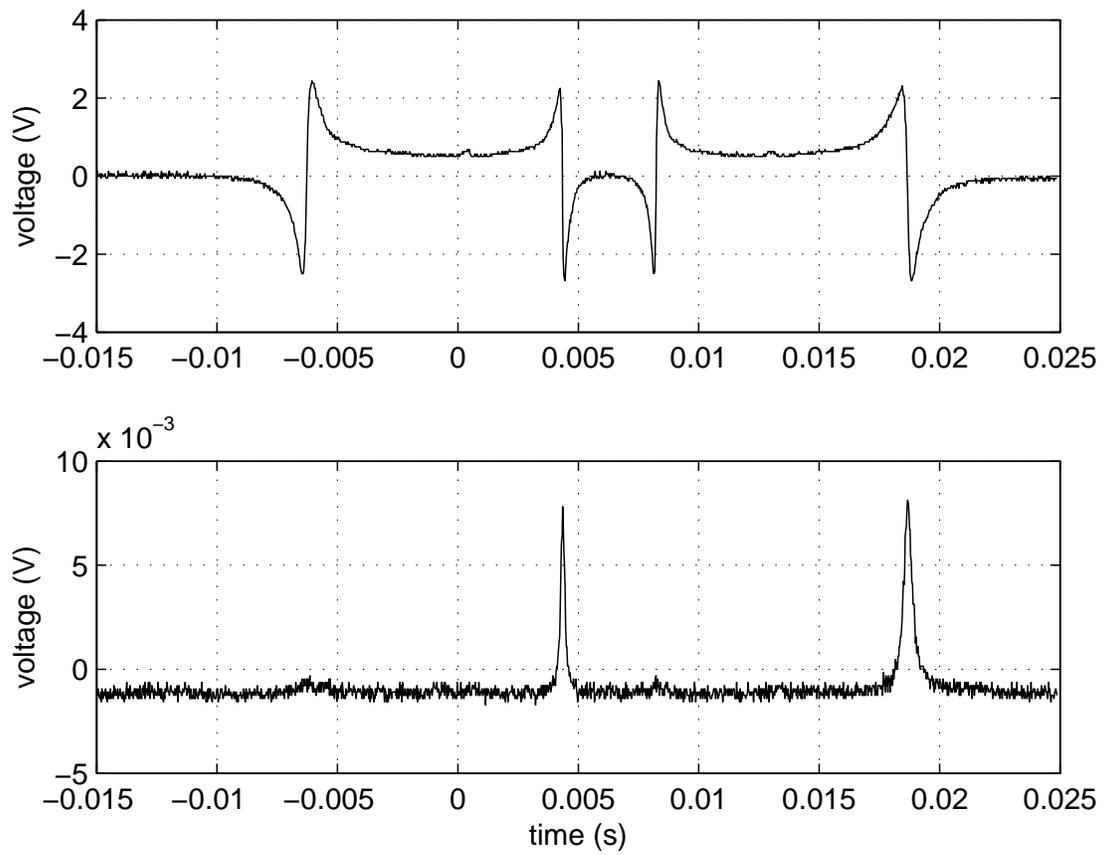}
  \caption{Error signal (top trace) and seed transmission through the OPO cavity (bottom trace) of the OPO cavity.}
  \label{error}
\end{figure}

The upper trace of Fig. \ref{error} shows a plot of $X_{sqz}^-$ measured using this technique.  In this polarization implementation, the seed and the local oscillator each produce a dispersion shaped error signal when resonant with the cavity, but with flipped sign relative to each other.  The lower trace in Fig. \ref{error} is a plot of the measured transmission of the seed field through the OPO cavity.  The error signal is used to lock the OPO cavity to a resonance for the seed mode, which corresponds to one of the peaks in the lower trace in Fig. \ref{error}.

\begin{figure}[htbp]
  \centering
  \vspace{-2pt}
  \includegraphics[width=\linewidth]{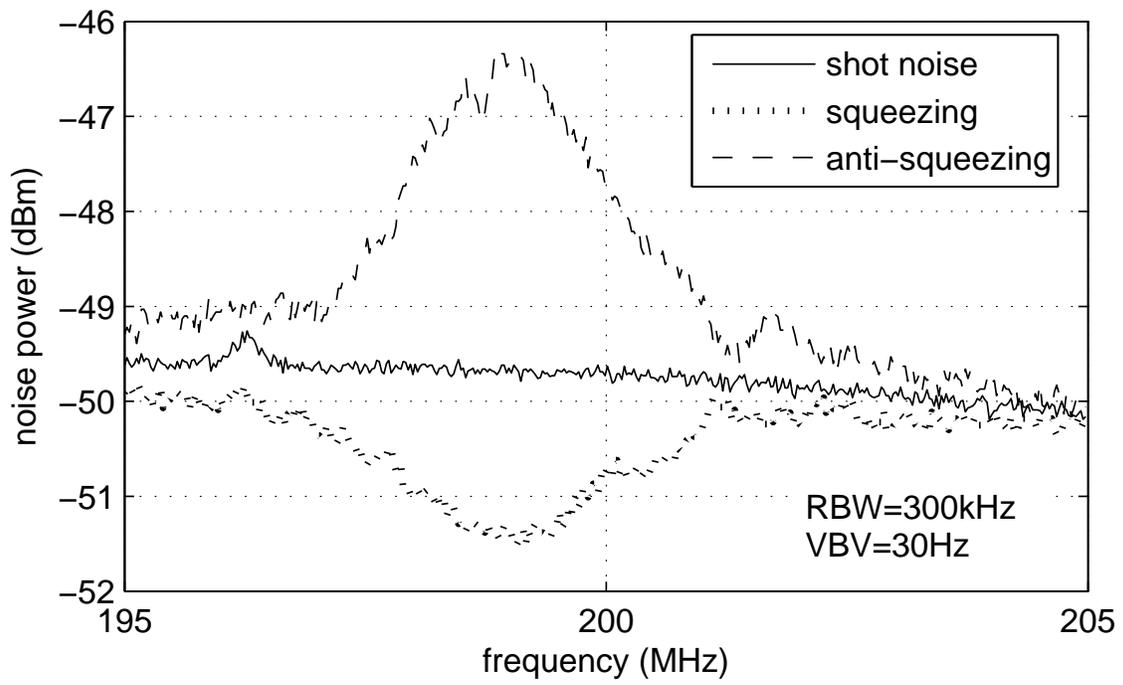}
  \caption{Squeezing spectrum at the first FSR of the OPO. }
  \label{sqz}
\end{figure}

The fluctuating component of the same homodyne measurement that lead to Fig. \ref{error} can be used to measure the squeezed output of the OPO.  Fig. \ref{sqz} is a plot of the variance of the homodyne measurement,  $|\delta X_{sqz}^-|^2$, at frequencies near the first
resonance of the OPO, made when the feedback controller is active.  Maximum squeezing of approximately 2dB has been observed at the resonant frequency
of 199MHz. Coupled with a measured de-amplification of 2.6dB, this measurement suggests that the overall escape and detection efficiency of the
system is greater than 87\% \cite{takeno07}.   

With thoughtful design, the homodyne locking technique can work even
with an experiment interposed between the OPO and the homodyne detector.  Furthermore, a co-propagating local oscillator turns out to be a most
useful addition to subsequent experiments.  

\begin{figure}[htbp]
  \centering
  \vspace{-2pt}
  \includegraphics[width=\linewidth,viewport=75 360 570 550,clip]{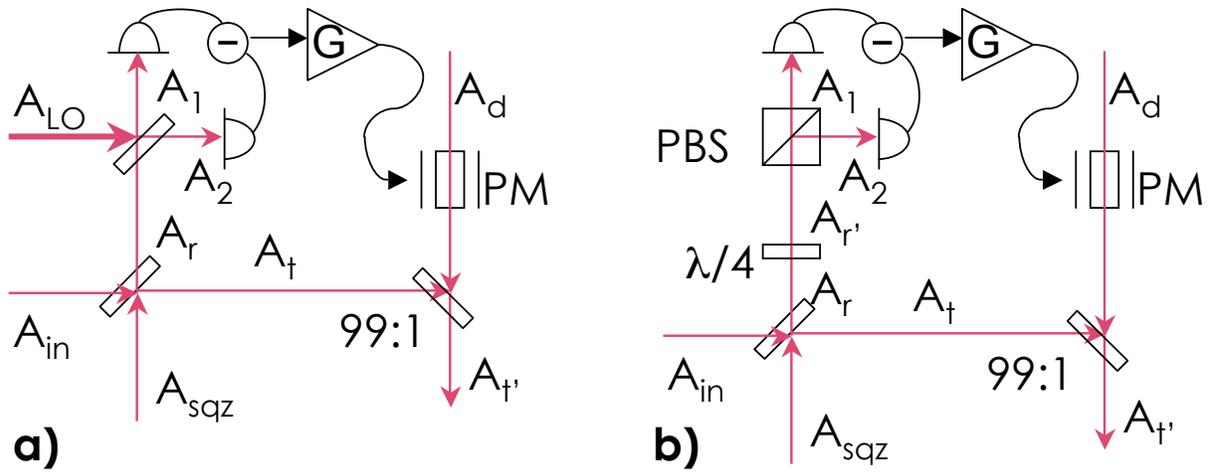}
  \caption{(a) Schematic of universal squeezer \cite{yoshikawa}. (b) Alternative implementation.}
  \label{uniSqz}
\end{figure}

For example, consider Fig. \ref{uniSqz}(a) in which the schematic diagram for a universal squeezer is reproduced from Ref. \cite{yoshikawa}.
Here, an input field is interferred with a phase-quadrature-squeezed vacuum on a beamsplitter of transmittivity $T$.  Phase quadrature detection
of one port of the beamsplitter is performed in the usual fashion \cite{BR04}.  The homodyne photocurrent is amplified by $g=-\sqrt{(1-T)/T}$ and fed forward to the other port of the beamsplitter.
After feed-forward, the output is a phase-quadrature squeezed version of the input, with excellent purity \cite{filip}.

Not shown in Fig. \ref{uniSqz}(a) are the requirements to lock the frequency of the OPO cavity and the phase of the LO.  Fig. \ref{uniSqz}(b) shows a schematic diagram for an alternative
implementation.  In that implementation, the homodyne locking technique can be used to simultaneously lock the frequency of the
OPO cavity and the LO phase.  The diagram is shown with polarisation modes, but the idea could
be equally well implemented using displaced beams, higher order spatial modes and so forth.  

Treating the input and output of the OPO cavity as vector quantities, $\tilde A_{c}=A_{s}\hat{x} + A_{LO}\hat{y}$ and $\tilde A_{sqz} = A_{sqz} \hat{x} + A_{LO}\hat{y}$, it is straightforward to trace the change in notation through Eqs. \ref{cavity-1}, \ref{alpha} and \ref{squeezingSimple}.  We treat $\hat{x}$ as being the desired cavity mode, and $\hat{y}$ as an orthogonal mode in one of the degrees of freedom.  The latter is not resonant with the OPO cavity, but it co-propagates with the former and interacts with all of the same linear optical elements.  The steady-state value of the homodyne photocurrent is $X_{out}^-$ as required to frequency-lock the OPO cavity and the linearised fluctuating component of the homodyne photocurrent is as required for the universal squeezing operation \cite{yoshikawa} when $\Delta=0$.   

In summary, we have proposed a new approach to controlling the resonant frequency of a sub-threshold OPO.  This approach is based on direct
measurement of the squeezed phase quadrature output of the OPO, rather than on the traditional approach where the squeezed output is not used for
frequency-locking.  The homodyne locking technique is cheap, easy to implement and, uniquely, is a sub-quantum-noise-limited frequency discriminator.  We have demonstrated the homodyne locking technique on a sub-threshold OPO producing approximately 2dB of squeezing.  We have
also shown, via an explicit example, that homodyne locking remains useful when an experiment is interposed between the OPO and the homodyne
detector.  Indeed this approach greatly simplifies the experiment as it relieves the necessity of locking subsequent homodyne detectors.

{\it Acknowledgments -} This work was supported by the Australian
Research Council.

\newpage

\section*{List of Figure Captions}
Fig. 1. Schematic of homodyne locking experiment.\\
Fig. 2. Error signal (top trace) and seed transmission through the OPO cavity (bottom trace) of the OPO cavity.\\
Fig. 3. Squeezing spectrum at the first FSR of the OPO. \\
Fig. 4. (a) Schematic of universal squeezer \cite{yoshikawa}. (b) Alternative implementation.

\newpage

\newpage

\newpage

\newpage

\end{document}